\documentclass[aps,pra,twocolumn,groupedaddress]{revtex4}
\usepackage{graphicx}
\usepackage[usenames]{color}
\usepackage{multirow}
\usepackage{amsmath}
\usepackage[latin1]{inputenc}
\usepackage{array}

\begin{document}

\title{Magnetic Linear Birefringence Measurements Using Pulsed Fields}

\author{P. Berceau$^{1}$}
\author{M. Fouch\'e$^{1,2,3}$}
\author{R. Battesti$^1$}
\author{C. Rizzo$^1$}
\email{carlo.rizzo@lncmi.cnrs.fr}

\affiliation{ $^{1}$Laboratoire National des Champs Magn\'etiques Intenses, (UPR 3228, CNRS-UPS-UJF-INSA),
31400 Toulouse, France.\\
$^2$Université de Toulouse, UPS, Laboratoire Collisions Agrégats Réactivité, IRSAMC, F-31062 Toulouse, France.\\
$^3$CNRS, UMR 5589, F-31062 Toulouse, France.}

\date{\today}

\begin{abstract}
In this paper we present the realization of further steps towards
the measurement of the magnetic birefringence of the vacuum using
pulsed fields. After describing our experiment, we report the
calibration of our apparatus using nitrogen gas and we discuss the
precision of our measurement giving a detailed error budget. Our
best present vacuum upper limit is $\Delta
n\,\leq\,5.0\,\times\,10^{-20}$\,T$^{-2}$ per 4\,ms acquisition
time. We finally discuss the improvements necessary to reach our
final goal.
\end{abstract}


\maketitle

\section{Introduction}

Experiments on the propagation of light in a transverse magnetic field date from the beginning of the 20$^{\mathrm{th}}$ century.
In 1901 Kerr \cite{Kerr1901} and in 1902 Majorana \cite{Majorana1902} discovered that linearly polarized light,
propagating in a medium in the presence of a transverse magnetic field, acquires an ellipticity. In the following years,
this linear magnetic birefringence was studied in detail by Cotton and Mouton \cite{CottonMouton} and it is known nowadays as the Cotton-Mouton effect.
It corresponds to an index of refraction $n_\parallel$ for light polarized parallel to the magnetic field $B$ that is different from the index of refraction $n_\bot$
for light polarized perpendicular to the magnetic field. For symmetry reasons, the difference between $n_\parallel$ and $n_\bot$ is proportional to $B^2$.
Thus, an incident linearly polarized light exits from the magnetic field region elliptically polarized.
For a uniform $B$ over an optical path $L$, the ellipticity is given by:
\begin{eqnarray}
\Psi = \pi\frac{L}{\lambda}\Delta n B^2 \sin 2 \theta,\label{Eq:Psi}
\end{eqnarray}
where $\lambda$ is the wavelength of light in vacuum, $\Delta n$\,=\,$n_\parallel$\,-\,$n_\bot$ at $B = 1$\,T and $\theta$ is the angle between light polarization and magnetic field.

The Cotton-Mouton effect exists in any medium and quantum electrodynamics predicts that magnetic linear birefringence exists also in vacuum. It has been shown
around 1970 \cite{Bialynicka1970,Adler1971} thanks to the effective Lagrangian established in 1935 and 1936 by Kochel, Euler and Heisenberg \cite{Euler1935,Heisenberg1936}.
At the lowers two orders in $\alpha$, the fine structure constant, $\Delta n$ can be written as:
\begin{equation}
\Delta n = \frac{2}{15}\frac{\alpha^2\hbar^3}{m_\mathrm{e}^4c^5\mu_0} \left(1+\frac{25}{4\pi}\alpha \right) [\mathrm{T}^{-2}], \label{Eq:Delta_n_cste_fond}
\end{equation}
where $\hbar$ is the Planck constant over $2\pi$, $m_\mathrm{e}$ is the electron mass, $c$ is the speed of light in vacuum, and $\mu_0$ is the magnetic constant.
The $\alpha^2$ term is given in Ref.\,\cite{Bialynicka1970}.
The $\alpha^3$ term has been first reported in Ref.\,\cite{Ritus1975} and it corresponds to the lowest order radiative correction. Its value is about $1.5\%$ of the $\alpha^{2}$ term.
Using the 2010 CODATA recommended values for the fundamental constants \cite{{CODATA2010}}, Eq.\,(\ref{Eq:Delta_n_cste_fond}) gives  $\Delta n=(4.031699\pm0.000002)\times 10^{-24} [\mathrm{T}^{-2}]$.

As we see, the error due to the uncertainty of fundamental constants is negligible compared to the error coming from the fact that only first order QED radiative correction has been calculated.
The QED $\alpha^4$ radiative correction should affect the fourth digit and the QED $\alpha^5$ radiative correction the sixth digit.
Thus, a measurement of $\Delta n$ up to a precision of a few ppm remains a pure QED test.

Experimentally, the measurement of the Cotton-Mouton effect is
usually very challenging especially in dilute matter, thus all the
more so in vacuum. Several groups have attempted to observe vacuum
magnetic birefringence \cite{QandA2007,PVLAS2008}, but this very
fundamental prediction still has not been experimentally confirmed.

Gas measurements date back to 1938 \cite{Rizzo1997} and the first
systematic work of Buckingham et al. was published in 1967
\cite{Buckingham1967}. Investigations concerned benzene, hydrogen,
nitrogen, nitrogen monoxide and oxygen at high pressures, and
ethane. Since 1967, many more papers concerning the effect in
gases have been published and Cotton-Mouton effect experiments
have been employed as sensitive probes of the electromagnetic
properties of molecules \cite{Rizzo1997}.

The measurement of the Cotton-Mouton effect in gases is not only
important to test quantum chemical predictions. It is a crucial
test for any apparatus which is dedicated to the search for vacuum
magnetic birefringence. Measurement of the Cotton-Mouton effect in
a gas is a milestone in the improvement of the sensitivity of such
an apparatus. Typically measurements of the linear magnetic
birefringence in nitrogen gas are used to calibrate a setup
\cite{QandA2007,PVLAS2008,BFRT1993}.

In the following we present magnetic linear birefringence
measurements performed in the framework of our ``Bir\'efringence
Magn\'etique du Vide'' (BMV) project. It is based on the use of
strong pulsed magnetic fields, which is a novelty as far as linear
magnetic birefringence is concerned, and on a very high finesse
Fabry-Perot cavity to increase the effect to be measured by
trapping the light in the magnetic field region. The use of pulsed
fields for such a kind of measurements has been first proposed in
Ref. \cite{RizzoEPL}. In principle, pulsed magnetic fields can be
as high as several tens of Tesla, which increases the signal, and
they are rapidly modulated which decreases the $1/f$- flicker
noise resulting in an increase of the signal to noise ratio. Both
advantages are supposed to compensate the loss of duty cycle since
only few pulses per hour are possible. A feasibility study, which
discusses most of the technical issues related to the use of
pulsed fields coupled to precision optics for magnetic linear
birefringence measurements, can be found in Ref.\,\cite{EPJD_BMV}.

In this paper we present the realization of further steps towards
the measurement of the magnetic birefringence of the vacuum using
pulsed fields. After describing our BMV experiment, we report the
calibration of our apparatus with nitrogen gas and we discuss the
precision of our measurement giving a detailed error budget.
Finally, present vacuum upper limit is reported and we discuss the
perspectives to reach our final goal.

\section{Experimental setup and signal analysis}

\subsection{Apparatus}

\begin{figure}[h]
\begin{center}
\includegraphics[width=8cm]{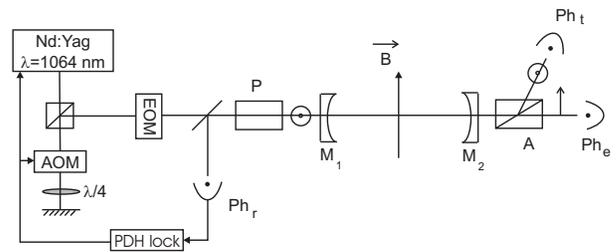}
\caption{\label{Fig:ExpSetup} Experimental setup. A Nd-YAG laser is
frequency locked to the Fabry-Perot cavity consisting of the mirrors M$_1$ and
M$_2$. The laser beam is linearly polarized by the polarizer P and
analyzed with the polarizer A. This analyzer allows to extract the
extraordinary beam sent on photodiode Ph$_\mathrm{e}$ as well as the
ordinary beam sent on photodiode Ph$_\mathrm{t}$. The beam reflected
by the cavity analyzed on the photodiode Ph$_\mathrm{r}$ is used for the cavity locking. A transverse
magnetic field $B$ can be applied inside the cavity in order to study the magnetic birefringence of the medium.
EOM = electro-optic modulator; AOM = acousto-optic modulator, PDH = Pound-Drever-Hall.}
\end{center}
\end{figure}

The BMV experiment is detailed in Ref.\,\cite{EPJD_BMV}. Briefly,
as shown on Fig.\,\ref{Fig:ExpSetup}, 30\,mW of a linearly
polarized Nd:YAG laser beam ($\lambda = 1064$\,nm) is injected
into a Fabry-Perot cavity consisting of the mirrors M$_1$ and
M$_2$. The laser frequency is locked to the cavity resonance
frequency using the Pound-Drever-Hall method \cite{PDH}. To this
end, the laser is phase-modulated at 10\,MHz with an electro-optic
modulator (EOM). The beam reflected by the cavity is then detected
by the photodiode Ph$_\mathrm{r}$. This signal is used to drive
the acousto-optic modulator (AOM) frequency for a fast control and
the Peltier element of the laser for a slow control of the laser
frequency.

Our birefringence measurement is based on an ellipticity
measurement. Light is polarized just before entering the cavity by
the polarizer P. The beam transmitted by the cavity is then analyzed
by the analyzer A crossed at maximum extinction and collected by a
low noise photodiode Ph$_\mathrm{e}$ (intensity of the extraordinary
beam $I_\mathrm{e}$). The analyzer also has an escape window which
allows us to extract the ordinary beam (intensity $I_\mathrm{t}$)
which corresponds to the polarization parallel to P. This beam is
collected by the photodiode Ph$_\mathrm{t}$.

All the optical components from the polarizer P to the analyzer A
are placed in a ultra high vacuum chamber. In order to perform
birefringence measurements on high purity gases, the vacuum
chamber is connected to several gas bottles through leak valves
which allow to precisely control the amount of injected gas.
Finally, since the goal of the experiment is to measure magnetic
birefringence, magnets surround the vacuum pipe. The transverse
magnetic field is created thanks to pulsed coils described in
Ref.\,\cite{Batut2008} and briefly detailed in the next section.

Both signals collected by the photodiodes outside the cavity are
simultaneously used in the data analysis as follows:
\begin{equation}
\frac{I_\mathrm{e}}{I_\mathrm{t}} = \sigma^2 + \Psi_\mathrm{tot}^2,
\end{equation}
where $\Psi_\mathrm{tot}$ is the total ellipticity acquired by the
beam going from P to A and $\sigma^2$ is the polarizer extinction
ratio. Our polarizers are Glan laser prisms which have an extinction
ratio of $2\times 10^{-7}$.

The origin of the total ellipticity  of the cavity is firstly due
to the intrinsic birefringence of the mirrors M$_1$ and M$_2$, as
it will be discussed in section\,\ref{subsubsec : birefringence}.
We define the ellipticity imparted to the linearly polarized laser
beam when light passes trough each mirror substrate as
$\Gamma_\mathrm{s1,2}$, and the one induced by the reflecting
layers of the mirrors as $\Gamma_\mathrm{c}$. An additional
component $\Psi$ of the total ellipticity can be induced by the
external magnetic field. Since we use pulsed magnetic fields, this
ellipticity is a function of time.

Finally, if the ellipticities are small compared to unity, one gets:
\begin{equation}
\frac{I_\mathrm{e}(t)}{I_\mathrm{t}(t)} = \sigma^2 + [\Gamma +
\Psi(t)]^2, \label{Eq:Ellipticitymeas}
\end{equation}
where $\Gamma = \Gamma_\mathrm{s1} + \Gamma_\mathrm{s2} +
\Gamma_\mathrm{c}$ is the total static birefringence.

\subsection{Magnetic field}

It is clear from Eq.\,(\ref{Eq:Psi}) that one of the critical
parameter  for experiments looking for magnetic birefringence is
$B^2L$. Our choice has been to reach a $B^2L$ as high as possible
having a $B$ as high as possible with a $L$ such as to set-up a
table-top low noise optical experiment. This is fulfilled using
pulsed magnets that can provide fields of several tens of Tesla. Our
apparatus consists of two magnets, called Xcoils. The principle of
these magnets and their properties are described in details in
Refs.\,\cite{Batut2008,EPJD_BMV}.

\begin{figure}[h]
\begin{center}
\includegraphics[width=8cm]{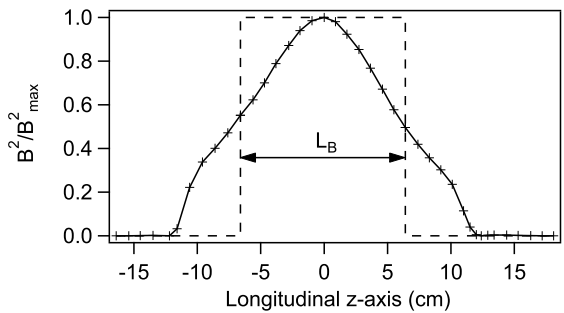}
\caption{\label{Fig:Profilefield} Normalized profile of the square
of the magnetic field along the longitudinal \textit{z-}axis
(solid line) inside one Xcoil. This is compared to the equivalent
uniform magnetic field (dashed line) over the effective magnetic
length L$_B$ (see text).}
\end{center}
\end{figure}

The magnetic field profile along the longitudinal \textit{z-}axis,
which corresponds to the axis of propagation of the light beam,
has been measured with a calibrated pick-up coil.
Fig.\,\ref{Fig:Profilefield} shows the normalized profile of an
Xcoil. The magnetic field is not uniform along \textit{z}. We
define $B_{\mathrm{max}}$ as the maximum field provided by the
coil at its center and $L_B$ as the equivalent length of a magnet
producing a uniform magnetic field $B_{\mathrm{max}}$ such that:
\begin{equation}
\int_{-\infty}^{+\infty}B^{2}(z)dz = B_{\mathrm{max}}^2 L_{B}.
\label{Eq:defLb}
\end{equation}
$L_B$ is about the half of the Xcoil's length. Each Xcoil
currently used has reached more than 14\,T over 0.13\,m of
effective length corresponding to 25 T$^2$m. The total duration of
a pulse is a few milliseconds. The magnetic field reaches its
maximum value within 2 ms.

The pulsed coils are immersed in a liquid nitrogen cryostat to
limit the consequences of heating which could be a cause of
permanent damage of the coil's copper wire. Pulse duration is
short enough that the coil, starting at liquid nitrogen
temperature, always remains at a safe level \textit{i.e.} below
room temperature. A pause between two pulses is necessary to let
the magnet cool down to the equilibrium temperature which is
monitored via the Xcoils' resistance. The maximum repetition rate
is 5 pulses per hour.

\subsection{Fabry-Perot cavity}

The other key point of our experiment is to accumulate the effect
due to the magnetic field by trapping the light between two ultra
high reflectivity mirrors constituting a Fabry-Perot cavity. Its
length has to be large enough to leave a wide space so as to
insert our two cylindrical cryostats (diameter of 60 cm for each
cryostat) and vacuum pumping system. The length of the cavity is
$L_\mathrm{c}=2.27$\,m corresponding to a free spectral range of
$FSR=c/2nL{_\mathrm{c}} \simeq $66\,MHz with $n$ the index of
refraction of the considered medium in which the cavity is
immersed. This index of refraction can be considered equal to one.
The total acquired ellipticity $\Psi$ is linked to the ellipticity
$\psi$ acquired in the absence of cavity, and depends on the
cavity finesse $F$ as follows \cite{brandi}:
\begin{equation}
\Psi = \frac{2F}{\pi}\psi, \label{Eq:linkPsi1Psitot}
\end{equation}
where $F$ is given by:
\begin{equation}
F = \frac{\pi \sqrt{R_{\mathrm{M}}}}{1-R_{\mathrm{M}}},
\end{equation}
with $R_\mathrm{M}$ the intensity reflection coefficient supposed to
be the same for both mirrors. In order to increase the induced
signal, a finesse as high as possible is essential.

\subsubsection{Cavity finesse and transmission}

\begin{figure}[h]
\begin{center}
\includegraphics[width=8cm]{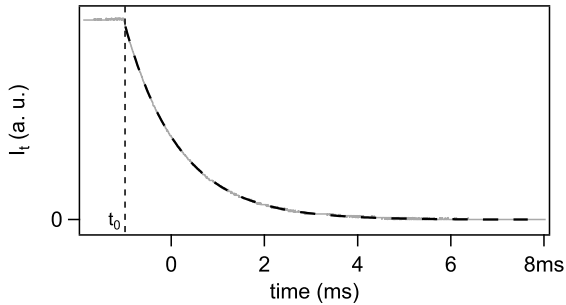}
\caption{\label{Fig:finesse_ATF} Time evolution of the intensity of
the ordinary beam (gray line). The laser is switched off at $t =
t_0$. Experimental data are fitted by an exponential decay (black
dashed line) giving a photon lifetime of $\tau$\,=\,1.16\,ms, a
finesse of $F\,=\,481\,000$ and a linewidth of $\Delta
\nu\,=\,c/2nL_\mathrm{c}F=137$\,Hz.}
\end{center}
\end{figure}

Experimentally, the finesse is inferred from a measurement of the
photon lifetime $\tau$ inside the cavity as presented on
Fig\,\ref{Fig:finesse_ATF}. For $t<t_0$, the laser is locked to the
cavity. The laser intensity is then switched off at $t_0$ thanks to
the AOM shown on Fig.\,\ref{Fig:ExpSetup} and used as an ultrafast
commutator. For $t>t_0$, one sees the typical exponential decay of
the intensity of the transmitted ordinary beam \cite{svelto}:
\begin{eqnarray}
I_\mathrm{t}(t)=I_\mathrm{t}(t_{0})e^{-(t-t_{0})/\tau}.\label{Eq:DureeVie_Iord}
\end{eqnarray}
The photon lifetime is related to the finesse of the cavity through
the relation:
\begin{equation}
\tau = \frac{n L_\mathrm{c} F}{\pi c}. \label{Eq:tau}
\end{equation}
By fitting our data with Eq.\,(\ref{Eq:DureeVie_Iord}) we get $\tau
=1.16$\,ms corresponding to a finesse of $F=481\,000$ and a cavity
linewidth of $\Delta \nu = c/2nL_\mathrm{c}F=137$\,Hz.

We summarize in Table \,\ref{Table:Interferometers} the performances
of some well-known sharp cavities at $\lambda = 1064$\,nm. To our
knowledge, currently our interferometer is the sharpest in the
world.

\begin{table*}
\center
\begin{tabular*}{1\textwidth}{@{\extracolsep{\fill}} c c r r r r r c }
    \hline
    \hline
    Interferometer & Ref.      & $L{_\mathrm{c}}$(m) & $FSR$(kHz)& $F$      & $\tau$($\mu$s) & $\Delta \nu$(Hz) & $Q$\\
    \hline
   VIRGO   & \cite{VIRGO}      & 3000     &      50   &       50 &  160 & 1000 & 2.8 $\times$ $10^{11}$ \\
   TAMA300 & \cite{TAMA}       &  300     &     500   &      500 &  160 & 1000 & 2.8 $\times$ $10^{11}$ \\
   PVLAS   & \cite{PVLAS2008}  &    6.4   &  23 400   &  70\,000 &  475 &  335 & 8.4 $\times$ $10^{11}$ \\
   LIGO    & \cite{LIGO}       & 4000     &      37   &      230 &  975 &  163 & 17  $\times$ $10^{11}$ \\
   BMV     & this work         &    2.27  &  66 000   & 481\,000 & 1160 &  137 & 21  $\times$ $10^{11}$ \\
    \hline
    \hline
 \end{tabular*}
\caption{\label{Table:Interferometers} Performances summary of the
sharpest infra-red interferometers in the world. $L{_\mathrm{c}}$ is
the length of the Fabry-Perot cavity, $FSR$ is its full spectral
range, $F$ is the cavity finesse, $\tau$ is the photon lifetime,
$\Delta \nu$ is the frequency linewidth and Q =
$\nu_{\mathrm{laser}}/\Delta \nu$ is the quality factor of the
interferometer, with $\nu_{\mathrm{laser}}$ the laser frequency.}
\end{table*}

The transmission of the cavity $T_\mathrm{c}$ is another important
parameter. It corresponds to the intensity transmitted by the cavity
divided by the intensity incident on the cavity when the laser
frequency is locked. Indeed in order not to be limited by the noise
of photodiodes Ph$_\mathrm{t}$ and Ph$_\mathrm{e}$, $I_\mathrm{t}$
and $I_\mathrm{e}$ have to be sufficiently high. This point is
particularly critical for $I_\mathrm{e}$ which corresponds to the
intensity transmitted by the cavity multiplied by $\sigma^2$. With a
Ph$_\mathrm{e}$ noise equivalent power of
11\,fW/$\sqrt{\mathrm{Hz}}$, we need an incident power greater than
0.2\,nW so as not to be limited by electronic noise of
Ph$_\mathrm{e}$.

Our cavity transmission is 20\,$\%$. The measurements of the finesse
and the transmission allow to calculate mirrors properties such as
their intensity transmission $T_\mathrm{M}$ and their losses
$P_\mathrm{M}$ thanks to the following relations:
\begin{eqnarray}
F &=& \frac{\pi}{T_\mathrm{M}+P_\mathrm{M}},\\
T_\mathrm{c} &=& \left(\frac{T_\mathrm{M}F}{\pi}\right)^2,
\end{eqnarray}
supposing that the mirrors are identical. We found $T_\mathrm{M} =
3$\,ppm and $P_\mathrm{M} = 3.5$\,ppm, which corresponds to the
specifications provided by the manufacturer.

To conclude, our high finesse cavity enhances the Cotton-Mouton
effect of a factor $2F/\pi$=306\,000, and its transmission allows
measurements that are not limited by the noise of the detection
photodiodes.

\subsubsection{Cavity birefringence}
\label{subsubsec : birefringence} The origin of the total static
ellipticity is due to the mirror intrinsic phase retardation.
Mirrors can be regarded as wave plates and for small
birefringence, combination of both wave plates gives a single wave
plate. The phase retardation and the axis orientation of this
equivalent wave plate depend on the birefringence of each mirror
and on their respective orientations \cite{Jacob, brandi}.

The intrinsic phase retardation of the mirrors is a source of noise
limiting the sensitivity of the apparatus. Moreover, since our
signal detection corresponds to a homodyne technique, the static
ellipticity $\Gamma$ is used as a zero frequency carrier. To reach a
shot noise limited sensitivity, one needs $\Gamma$ to be as small as
possible \cite{EPJD_BMV}, implying that the phase retardation axis
of both mirrors have to be aligned. For magnetic birefringence
measurements, both mirrors' orientation is adjusted in order to have
$10^{-3}< \Gamma < 3\times 10^{-3}$\,rad.

Measurement of the total ellipticity as a function of mirror
orientation allows to calculate the mirror intrinsic phase
retardation per reflection. The experimental procedure is
presented in Ref.\,\cite{BirMirror}. The deduced phase retardation
for our mirrors is $\delta_\mathrm{M} = (7 \pm 6)\times
10^{-7}$\,rad. Although the origin of the mirrors' static
birefringence is still unknown, a review of existing data shows
that for interferential mirrors phase retardation per reflection
decreases when reflectivity increases \cite{BirMirror}. This
observation is confirmed by our new measurement. It is also in
agrement with the empirical trend given in Ref.\,\cite{BirMirror}:
$\delta_\mathrm{M}\simeq 0.1\times(1-R_{\mathrm{M}})$. Numerical
calculations show that this trend can be explained assuming that
the effect is essentially due to the layers close to the
substrate.

As said before, mirror birefringence has two contributions: one
comes from the substrate while the other is due to the reflecting
layers. Whereas previous measurements do not allow to distinguish
between the two contributions, we will see that this can be achieved
with the measurement of $I_\mathrm{e}$ decay.

\begin{figure}[h]
\begin{center}
\includegraphics[width=8cm]{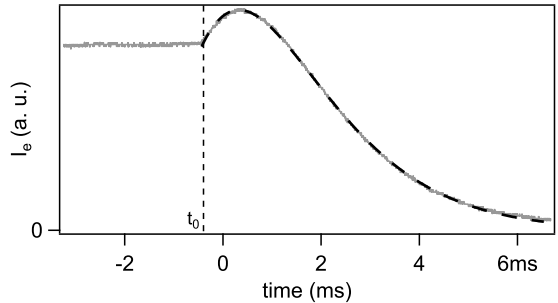}
\caption{\label{Fig:finesse_Ie} Time evolution of the intensity of
the extraordinary beam (gray line). The laser is switched off at $t
= t_0$. Experimental data are perfectly fitted by
Eq.\,(\ref{Eq:DurreVie_Iext}) (black dashed line).}
\end{center}
\end{figure}

A typical time evolution of $I_\mathrm{e}$ when the incident beam
locked to the cavity is switched off is shown in
Fig.\ref{Fig:finesse_Ie}. We see that this curve can not be fitted
by an exponential decay. As explained in Ref.\,\cite{Berceau2010},
one has to take into account the intrinsic birefringence of the
cavity. Nevertheless, the expression derived in
Ref.\,\cite{Berceau2010}, which only takes into account the
reflecting layers birefringence, does not always fit our data. The
evolution of $I_\mathrm{e}$ presents sometimes an unexpected
behavior: whereas no photon enters anymore into the cavity at
$t=t_0$, the extraordinary intensity starts growing before
decreasing. To reproduce this behavior, one has to take into account
the substrate birefringence.

Lets calculate the transmitted intensity along the round-trip inside
the cavity:
\begin{itemize}
\item For $t \leq t_{0}$, the laser is continuously locked to the
cavity. According to Eq.\,(\ref{Eq:Ellipticitymeas}), the
intensities of the ordinary and the extraordinary beams are related
by:
\begin{eqnarray}
I_\mathrm{e}(t\leq t_{0})=[\sigma^2 + (\Gamma_\mathrm{s2} +
\Gamma_\mathrm{s1} + \Gamma_\mathrm{c})^{2}]I_\mathrm{t}(t\leq
t_{0}).
\end{eqnarray}

\item At $t = t_{0}$, the laser beam is abruptly switched off, the cavity
empties gradually. The ordinary and extraordinary beams are slightly
transmitted at each reflection on the mirrors. But, because these
mirrors are birefringent, some photons of the ordinary beam are
converted into the extraordinary one. The reverse effect is
neglected because $I_\mathrm{e}\ll I_\mathrm{t}$.
\end{itemize}

We then follow the same procedure as in Ref.\,\cite{Berceau2010} to
calculate the time evolution of $I_\mathrm{e}$. For $t>t_0$, one
gets:
\begin{eqnarray}
  \lefteqn{I_{e}(t) = } \label{Eq:DurreVie_Iext} \\
  & & I_{t}(t) \left(\sigma^2 + \left[\Gamma_\mathrm{s1} + \Gamma_\mathrm{s2} + \Gamma_\mathrm{c} \left(1+\frac{t-t_0}{2\tau}\right) \right]^{2}\right).  \nonumber
\end{eqnarray}
The behavior shown on Fig.\,\ref{Fig:finesse_Ie} is reproduced if
$(\Gamma_\mathrm{s1}+\Gamma_\mathrm{s2})\,\simeq\,
-\Gamma_\mathrm{c}$.

This expression is used to fit our experimental data plotted on
Fig.\,\ref{Fig:finesse_Ie}. We find a photon lifetime of $\tau =
960\,\mu$s which is in good agreement when fitting $I_\mathrm{t}$
\cite{It_Ie_justification}, $(\Gamma_\mathrm{s1}+\Gamma_\mathrm{s2})
= 2\times10^{-3}$\,rad and $\Gamma_\mathrm{c} =
-7\times10^{-3}$\,rad.

We have for the first time the evidence that the substrate is
birefringent and that this birefringence contributes to the total
ellipticity due to the cavity.

\subsection{Signal analysis}
The starting point of our analysis are the voltage signals
$V_\mathrm{e}$ and $V_\mathrm{t}$ provided by Ph$_\mathrm{e}$ and
Ph$_\mathrm{t}$. Voltage signals have to be converted into intensity
signals by using the photodiode conversion factor $g_\mathrm{e}$ and
$g_\mathrm{t}$:
\begin{eqnarray}
I_{\mathrm{e}} = g_\mathrm{e} V_\mathrm{e},\\
I_{\mathrm{t}} = g_\mathrm{t} V_\mathrm{t}.
\end{eqnarray}
As demonstrated in Ref.\,\cite{Berceau2010}, before analyzing raw
signals one has to take into account the first order low pass
filtering of the cavity. $I_{\mathrm{t,filtered}}$ in the Fourier
space is given by:
\begin{equation}
I_{\mathrm{t,filtered}}(\omega) =
\frac{1}{1+i\frac{\omega}{\omega_\mathrm{c}}}
I_{\mathrm{t}}(\omega), \label{Eq:filterIt}
\end{equation}
where $\nu_\mathrm{c}=\omega_\mathrm{c}/2 \pi=1/4 \pi \tau$ is the
cavity cutoff frequency. Then, according to
Eq.\,(\ref{Eq:Ellipticitymeas}), the ellipticity $\Psi(t)$ to be
measured can be written as:
\begin{equation}
\Psi(t) = -\Gamma +
\sqrt{\frac{I_\mathrm{e}(t)}{I_\mathrm{t,filtered}(t)}-\sigma^2}.
\label{Eq:Psi_t}
\end{equation}
The total static birefringence $\Gamma$ is measured a few
milliseconds just before the beginning of the magnetic pulse, thus
when $\Psi(t) = 0$.

On the other hand, $\Psi$ is proportional to the square of the
magnetic field and thus can be written as:
\begin{eqnarray}
\Psi(t) = \kappa B^2_\mathrm{filtered}(t). \label{Eq:otherPsi}
\end{eqnarray}
Since the photon lifetime is comparable with the rise time of the
magnetic field, the first-order low pass filtering of the cavity has
also to be taken into account on the quantity $B^2(t)$ as in
Ref.\,\cite{Berceau2010}. To recover the value of the constant
$\kappa$ we calculate for each pulse the correlation between
$\Psi(t)$ and $B^2_\mathrm{filtered}(t)$:
\begin{equation}
\kappa = \frac{\int_0^{T_{\mathrm{i}}}
\Psi(t)B(t)_\mathrm{filtered}^2 dt}{\int_0^{T_{\mathrm{i}}}
[B(t)^2_\mathrm{filtered}]^2 dt}, \label{Eq:kappa}
\end{equation}
where $T_\mathrm{i}$ is the integration time. A statistical analysis
gives the mean value of $\kappa$ and its uncertainty.

The magnetic birefringence $\Delta n$ is finally given by:
\begin{equation}
\Delta n (T,P) = \frac{\kappa}{4\pi\,\tau\,
FSR}\times\frac{\lambda}{L_B} \times \frac{1}{\sin 2\theta}.\\
\label{Eq:Ellipticitygeneral}
\end{equation}
$\Delta n$ is thus expressed in T$^{-2}$. $T$ and $P$ correspond to
gas temperature and pressure when measurements of magnetic
birefringence on gases are performed. We define the normalized
birefringence $\Delta n_\mathrm{u}$ as $\Delta n$ for $P = 1$\,atm
and $B=1$\,T.

\section{Experimental parameters and error budget}
In the following, to evaluate the precision of our apparatus in the
present version, we list the uncertainties at 1$\sigma$ on the
measurement of the parameters of Eq.\,(\ref{Eq:Ellipticitygeneral})
as recommended in Ref.\,\cite{CODATA1998}. The uncertainty on the
magnetic birefringence has two origins. The evaluation of the
uncertainty by a statistical analysis of series of observations is
termed a type A evaluation and mainly concerns the measurement of
$\tau$ and $\kappa$. An evaluation by means other than the
statistical analysis of series of observations, calibrations for
instance, is termed a type B evaluation and especially affects the
parameters $B$, $FSR$, $L_B$, $\lambda$ and $\theta$.

\subsection{Photon lifetime in the Fabry-Perot Cavity}
The photon lifetime $\tau$ is measured by analyzing the
exponential decay of the intensity of the transmitted light.
Several measurements have been performed both before and after
almost each magnetic pulse. The uncertainty on the value of $\tau$
comes from the fact that mirrors can slightly move because of
thermal fluctuations and acoustic vibrations. Measurements
conducted in the same experimental conditions have been studied
statistically leading to a relative variation of $\tau$ that does
not exceed 2\,\% at 1$\sigma$-level. Data taken during operation,
\textit{i.e.} before and after magnetic pulses, show the same
statistical properties as the ones taken without any magnetic
field. Thus, the magnetic field does not cause additional change
in $\tau$.

\subsection{Correlation factor}
\label{subsec : kappa}

The correlation factor $\kappa$ is given by Eq.\,(\ref{Eq:kappa}).
The A-type uncertainty on $\kappa$ depends on the measurement of
$\Psi$ and thus on the experimental parameters given in
Eq.\,(\ref{Eq:Psi_t}). In practice, we pulse the magnets several
times in the same experimental conditions to obtain a set of
values of $\kappa$. The distribution of the $\kappa$ values is
found to be gaussian, and we assume that its standard deviation
corresponds to the A-type uncertainty on $\kappa$. For our
measurements performed with nitrogen and presented in
section\,\ref{subsec : Apparatus calibration}, the A-type relative
uncertainty is typically 3.5\,$\%$. The standard uncertainty of
the average value of $\kappa$ can then be reduced increasing the
number of pulses.

B-type uncertainties depend on those of the square of the magnetic
field, the photodiode conversion factors, and the filter function
applied to the field.

To measure the magnetic field during operation, we measure the
current which is injected in our X-coil. As mentioned in
Ref.\,\cite{Batut2008}, the form factor $B/I$ has been determined
experimentally during the test phase by varying the current inside
the X-coil (modulated at room temperature or pulsed at liquid
nitrogen temperature), and by measuring the magnetic field induced
on a calibrated pick-up coil. These measurements have led to a
relative B-type uncertainty of $\delta B /B$ = 0.7\,\% for the
magnetic field corresponding to a B-type uncertainty on $\kappa$ of
1.4\,\%.

The ratio $g_\mathrm{e}/g_\mathrm{t}$ is measured from time to time
by sending the same light intensities to each photodiode. The
relative uncertainty in this parameter is $1.5$\,$\%$ corresponding
to the same amount relative uncertainty in $\kappa$.

$I_\mathrm{t}$(t) and $B^2$(t) are also filtered by a function that
involves the parameter $\tau$. We have empirically determined that a
$\tau$-variation of 2\% led to a $\kappa$-variation of 0.8\,\%.

We can finally add quadratically the uncertainties above, and
deduce that a B-type uncertainty of 2.2\,\% must be taken into
account on every measurement of the correlation factor $\kappa$.

\subsection{Frequency splitting between perpendicular polarizations}
In this section we evaluate the attenuation of the extraordinary
beam transmitted by our sharp resonant Fabry-Perot cavity on which
the laser's ordinary beam is frequency-locked. Let's suppose that
the ordinary (resp. extraordinary) beam resonates in the
interferometer at the frequency $\nu_\mathrm{t}$ (resp.
$\nu_\mathrm{e}$). The laser is locked to the cavity thanks to the
ordinary beam. Thus $\nu_\mathrm{t}$ corresponds to the top of the
transmission Airy function $\emph{A}$ of the Fabry-Perot cavity
which is given by:
\begin{equation}
\emph{A}(\nu) =
\frac{T_\mathrm{c}}{1+\frac{4F^{2}}{\pi^{2}}\sin^{2}\big(\frac{2\pi
n L_\mathrm{c}}{c}\nu\big)}. \label{airy}
\end{equation}

\begin{figure}[h]
\begin{center}
\includegraphics[width=8cm]{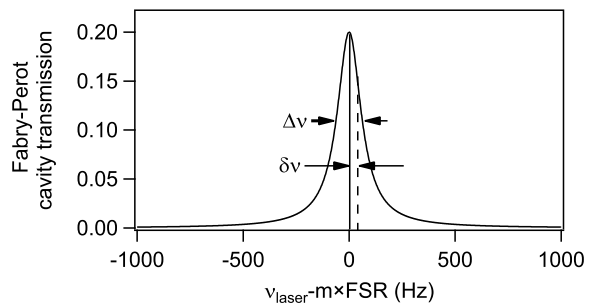}
\caption{\label{Fig:airy_deltanu} Airy function of our Fabry-Perot
cavity (linewidth $\Delta \nu$ = 137 Hz and transmission
$T_\mathrm{c}$\,=\,20\,\%). The frequency of the ordinary beam is
assumed to be locked at the top of the transmission function (solid
line) while he frequency $\nu_\mathrm{e}$ of the extraordinary beam
is shifted from $\nu_\mathrm{t}$ of a quantity $\delta \nu$ (dashed
line).}
\end{center}
\end{figure}

The frequency $\nu_\mathrm{e}$ is shifted from $\nu_\mathrm{t}$ by a
quantity $\delta \nu$ as it is shown on
Fig.\,\ref{Fig:airy_deltanu}.

The frequency splitting $\delta \nu = \nu_\mathrm{t}-\nu_\mathrm{e}$
can be expressed as a function of the phase retardation $\delta$
acquired along a round-trip between the ordinary and the
extraordinary beams:
\begin{eqnarray}
\delta \nu & = & \frac{c}{2 \pi n L_\mathrm{c}}\delta, \nonumber \\
& = & \frac{F\Delta \nu}{\pi} \delta.\label{deltanudelta}
\end{eqnarray}
This formula indicates that in order to have a splitting very small
compared to the cavity linewidth ($\delta\nu\,\ll\,\Delta\nu$), the
phase retardation $\delta$ must satisfy the following condition:
\begin{eqnarray}
\delta  \ll  \frac{\pi}{F},\label{Eq:Rayleighdelta}
\end{eqnarray}
which is equivalent to the condition on the acquired total
ellipticity $\Psi$:
\begin{eqnarray}
\Psi  \ll  1. \label{Eq:RayleighPsi}
\end{eqnarray}

\begin{figure}[h]
\begin{center}
\includegraphics[width=8cm]{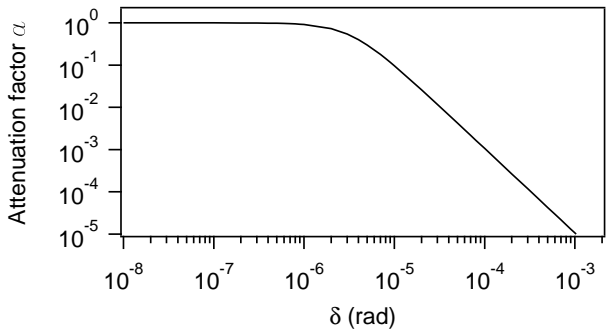}
\caption{\label{attenuationairy} Attenuation factor $\emph{a}$ as a
function of the phase retardation $\delta$ between both
polarizations.}
\end{center}
\end{figure}

By combining Eqs.\,(\ref{airy}) and (\ref{deltanudelta}), we obtain
the factor of attenuation $\emph{a}$ of the transmitted
extraordinary beam's intensity given by:
\begin{eqnarray}
\emph{a}& = &
\frac{\emph{A}(\nu_\mathrm{e})}{\emph{A}(\nu_\mathrm{t})},
\nonumber\\
& = & \frac{1}{1+\frac{4F^{2}}{\pi^{2}}\sin^{2}\big(\frac{2\pi n L_\mathrm{c}}{c}\delta \nu\big)}, \nonumber \\
& = & \frac{1}{1+\frac{4F^{2}}{\pi^{2}}\sin^{2}\big(\delta \big)}.
\end{eqnarray}
The attenuation factor $\emph{a}$ is plotted as a function of
$\delta$ on Fig.\,\ref{attenuationairy} for a finesse
$F$\,=\,481\,000. The \textit{real} intensity $I_\mathrm{e}$ of the
extraordinary beam transmitted by the cavity is obtained from the
corrected \textit{measured} intensity $I_\mathrm{e}^\mathrm{meas}$
as $I_\mathrm{e}\,=\,I_\mathrm{e}^\mathrm{meas}\,/\,\emph{a}$.

The frequency splitting can first be due to our birefringent cavity.
As in Ref.\,\cite{brandi}, let's consider both cavity mirrors
equivalent to a single wave-plate with phase retardation $\delta
_\mathrm{w}=\delta$ between both polarizations. The total phase
retardation $\delta_\mathrm{w}$ is linked to the cavity mirrors'
M$_1$ and M$_2$ own phase retardation $\delta_1$ and $\delta_2$ as
\cite{brandi}:
\begin{equation}
\delta_\mathrm{w}=
\sqrt{(\delta_1-\delta_2)^2+4\delta_1\delta_2\cos^2(\theta_\mathrm{m})}.
\end{equation}
To set a $\delta_\mathrm{w}$ as small as possible so as to minimize
the correction to $I_\mathrm{e}^\mathrm{meas}$, one needs to adjust
the angle $\theta_\mathrm{m}$ between the neutral axes of both
mirrors. This way, we set a $\delta_\mathrm{w}$ of the order of a
few $10^{-8}$ rad, corresponding to a correction smaller than
0.001\,\% on $I_\mathrm{e}^\mathrm{meas}$.

Secondly, the frequency splitting between both polarizations can be
due to the induced magnetic birefringence of the medium inside the
chamber. As seen above, the induced ellipticity given by
Eq.\,(\ref{Eq:RayleighPsi}) must be well below 1\,rad. This
condition is always satisfied in the range of pressure and field we
are working. The induced ellipticity does not exceed $10^{-2}$ rad.
This corresponds at worst to a phase retardation of $\delta$ =
$10^{-7}$ rad. The attenuation factor $I_\mathrm{e}^\mathrm{meas}$
is thus smaller than 0.1 $\%$.

In principle, this attenuation generates an error that has to be
taken into account in the measured ratio $I_\mathrm{e}/I_\mathrm{t,
filtered}$ of Eq.\,(\ref{Eq:Psi_t}), which implies an error in the
value of $\kappa$. At present, since the attenuation is smaller than
0.1\,\%, this error can be neglected compared to the others
uncertainties in $\kappa$.

\subsection{Cavity free spectral range}
\begin{figure}[h]
\begin{center}
\includegraphics[width=8cm]{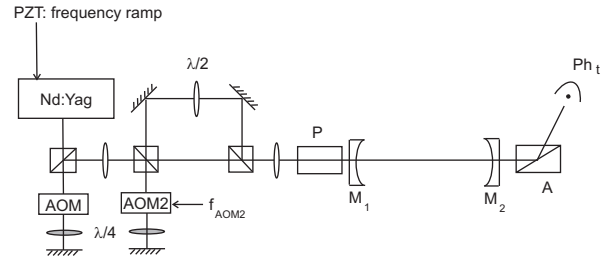}
\caption{\label{Fig:ExpSetupLcav} Experimental setup for the
cavity length measurement. Two laser beams  frequency shifted one
compared to the other by AOM2 are injected into the cavity. The
frequency of the laser is frequency modulated with a voltage ramp
applied on a piezo element mounted on the crystal resonator of the
laser. Photodiode Ph$_\mathrm{t}$ allows to observe the typical
Fabry-Perot peaks from which the $FSR$ measurement is performed.}
\end{center}
\end{figure}

The dedicated experimental setup for the measurement of the cavity
free spectral range $FSR = c/2 n L_\mathrm{c}$ is shown on
Fig.\,\ref{Fig:ExpSetupLcav}. The principle is to inject into the
cavity two laser beams shifted one compared with the other by a
given frequency. This frequency is then adjusted to coincide with
the free spectral range.

Experimentally, the main beam is divided into two parts thanks to a
polarizing beam splitting cube. The first part is directly injected
into the cavity while the other one is frequency shifted by the
acousto-optic modulator AOM2 with a double-pass configuration before
injection. The main beam is frequency modulated with a voltage ramp
applied on a piezo element mounted on the crystal resonator of the
laser.

\begin{figure}[h]
\begin{center}
\includegraphics[width=8cm]{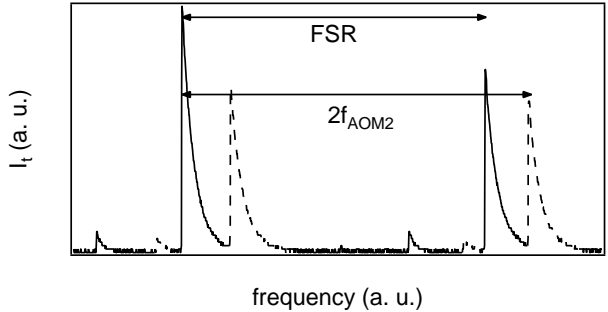}
\caption{\label{Fig:FSRPeaks} Transmission peaks of the Fabry-Perot
cavity as a function of the laser frequency. Two beams are sent to
the interferometer: the second beam (dashed line) is frequency
shifted by $2f_\mathrm{AOM2}$ compared to the first beam (solid
line). The adjustment of $f_{\mathrm{AOM2}}$ in order to superimpose
both series of peaks allows to precisely measure the free spectral
range $FSR$ of the cavity.}
\end{center}
\end{figure}

The intensity transmitted by the cavity is observed on
Ph$_\mathrm{t}$ as shown on Fig.\,\ref{Fig:FSRPeaks}. The solid line
corresponds to the intensity of the first beam. We observe typical
Fabry-Perot peaks whose frequency gap corresponds to $FSR$. Peaks
due to the second beam (dashed line) are frequency shifted by
$2f_{\mathrm{AOM2}}$. We finally adjust $f_{\mathrm{AOM2}}$ in order
to superimpose both series of peaks. The precise knowledge of the
driven frequency $f_\mathrm{AOM2}$ enables us to determine with the
same precision the value of the free spectral range, and thus the
cavity length.

A typical value is $FSR = (65.996 \pm 0.017)$\,MHz. This corresponds
to a cavity length of $L_\mathrm{c}=(2.2713\pm 0.0006)$\,m. Since
this length can be prone to variation, the $FSR$ value is regularly
checked and updated.

\subsection{Effective magnetic length}
Following Eq.\,(\ref{Eq:defLb}), the effective magnetic length
$L_B$ has been calculated by numerically integrating the field
measured with a calibrated pick-up coil. Taking into account the
experimental uncertainties, we got for one Xcoil:
$L_B$\,=\,(0.137\,$\pm$\,0.003)\,m, corresponding to a relative
B-type uncertainty on $L_{B}$ of 2.2\,\%.

\subsection{Laser wavelength}
As mentioned above, infra-red light enters the cavity. The
wavelength of the Nd:YAG laser is 1064\,nm, and its uncertainty is
given by the width of the laser transition. The natural linewidth
of Nd:YAG lasers are not usually given by the manufacturers.
However, we can estimate it from the bandwidth of the gain curve
of the amplifying medium. It is typically of the order of 30 GHz
\cite{Hecht}. This corresponds to an uncertainty on the laser
wavelength of 0.3 nm. In order to be conservative, we use $\lambda
= (1064.0\pm 0.5)$\,nm. The relative uncertainty is negligible in
our case, compared to main uncertainties.

\subsection{Angle between the incident polarization and the magnetic field direction}
The angle between the incident light polarization and the magnetic
field direction is adjusted to $45°$ thanks to magnetic
birefringence measurements as a function of the polarizer direction
$\theta_\mathrm{P}$. In order to be more sensitive, this is
performed close to the position where the magnetic field is parallel
to the polarizer P ($\theta = 0°$).

\begin{figure}[h]
\begin{center}
\includegraphics[width=8cm]{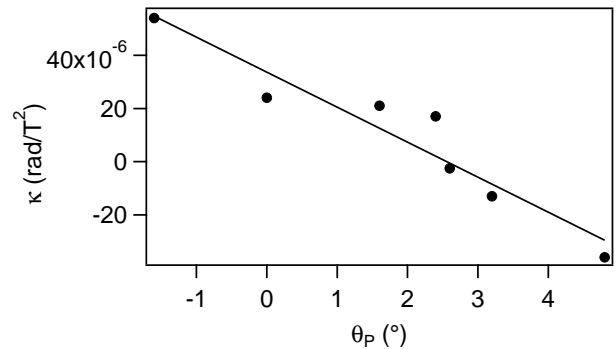}
\caption{\label{Fig:teta} Correlation factor $\kappa$ between the
square of the magnetic field and the ellipticity as a function of
the angle $\theta_{\mathrm{P}}$ of the incident polarization.}
\end{center}
\end{figure}

Measurements are realized with about 7$\,\times\,10^{-3}$\,atm of
air. The analyzer direction is crossed at maximum extinction each
time the polarizer is turned. Fig.\,\ref{Fig:teta} represents the
evolution of the correlation factor $\kappa$ as a function of
$\theta_{\mathrm{P}}$. Data are fitted by a sinusoidal trend $\kappa
(\theta_{\mathrm{P}})=\kappa_{0} \sin \bigl[
2(\theta_{\mathrm{P}}-\theta_{0}) \bigr]$ giving $\theta_{0}=(2.6
\pm 0.2)°$. This measurement allows to set $\theta$ to
$(45.0\pm1.2)°$. The uncertainty is mainly due to the mechanical
system which holds and turns the polarizer.

\subsection{Error budget}
We summarize in the Table \ref{TableBudget} the typical values of
the experimental parameters that have to be measured and their
B-type associated uncertainty. These uncertainties are quadratically
added to give a B-type relative uncertainty on the birefringence
$\Delta n$ of 3.1\,\% at 1$\sigma$.

\begin{table}
\begin{center}
\begin{tabular*}{0.48\textwidth}{p{2cm}  p{3cm}  r}
  \hline
  \hline
     \centering \multirow{2}{*} {Parameter} &   \centering Typical  &  Relative \\
 &  \centering value & B-type uncertainty\\
 \hline
 \centering $\kappa$         &  \centering $10^{-5}$\,rad\,T$^{-2}$     &  $2.2 \times 10^{-2}$ \\
 \centering $FSR$            &  \centering 65.996\,MHz &  $3 \times 10^{-4}$   \\
 \centering $L_{B}$          &  \centering 0.137\,m &  $2.2 \times 10^{-2}$   \\
 \centering $\lambda$        &  \centering 1064.0\,nm &   $<$ $5 \times 10^{-4}$   \\
 \centering $\sin 2\theta$   &  \centering 1.0000 &  $9 \times 10^{-4}$   \\
 \hline
 \multicolumn{2}{c}{total}&  $3.1 \times 10^{-2}$ \\
 \hline
 \hline
\end{tabular*}
\end{center}
\caption{\label{TableBudget} Parameters that have to be measured to
infer the value of the birefringence $\Delta n$ and their respective
relative B-type uncertainty at 1$\sigma$.}
\end{table}

\subsection{Temperature and pressure of gases}
Gas magnetic birefringence measurements are performed at room
temperature $T\,$=\,293\,K. The experimental room is
air-conditioned. A flow of compressed air between the outer wall of
the vacuum pipe and the liquid nitrogen cryostat containing the
magnet maintains the room temperature in the gas chamber.

\begin{figure}[h]
\begin{center}
\includegraphics[width=8cm]{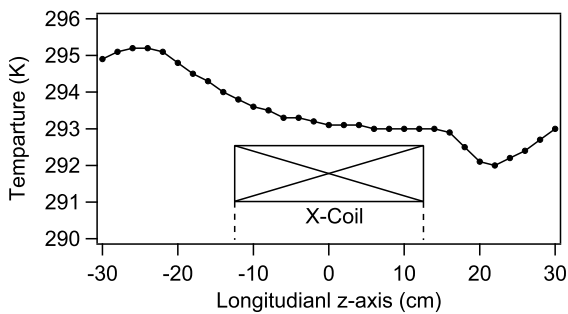}
\caption{\label{Fig:ProfileTemp} Profile of the temperature inside
the vacuum pipe along the longitudinal \textit{z-}axis. The X-coil
is also schematized at the center. The temperature variation does
not exceed 1\,K inside the tube that runs through it.}
\end{center}
\end{figure}

A temperature profile has been realized along the length of the
vacuum pipe, and is plotted on Fig.\,\ref{Fig:ProfileTemp}. The
temperature variation does not exceed 1\,K inside the tube that
passes through the magnetic field. Concerning gases, we consider
that our birefringence measurements are given at (293\,$\pm$\,1)\,K.

The pressure of the gas inside the chamber is measured at each
side of the vacuum pipe getting into magnets with pressure gauges.
The relative uncertainty provided by the manufacturer is 0.2\,\%.

\section{Magnetic birefringence measurements}

\subsection{Raw signals}
\begin{figure}[h]
\begin{center}
\includegraphics[width=8cm]{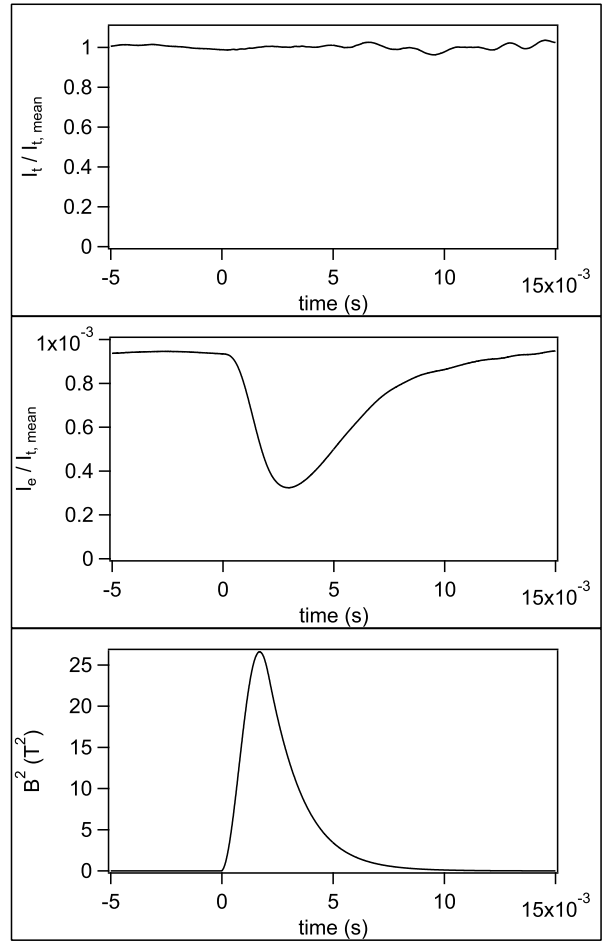}
\caption{\label{Fig:It_Ie_B2} Cotton-Mouton effect measurement on
32.1$\,\times\,10^{-3}$\,atm of molecular nitrogen. (top)
Normalized intensity of the ordinary beam as a function of time.
(middle) Intensity of the extraordinary beam divided by the mean
of $I_\mathrm{t}$ as a function of time. (bottom) Square of the
magnetic field as a function of time.}
\end{center}
\end{figure}

Fig.\,\ref{Fig:It_Ie_B2} presents signals obtained with
32.1$\,\times\,10^{-3}$\,atm of molecular nitrogen. The intensity
of the ordinary beam $I_\mathrm{t}$ (top) remains almost constant
while the intensity of the extraordinary beam $I_\mathrm{e}$
(middle) varies when the magnetic field (bottom) is applied. The
magnetic field reaches its maximum of 5.2\,T within less than
2\,ms.

The laser beam remains locked to the Fabry-Perot cavity, despite
mechanical vibrations caused by the shot of magnetic field.
$I_\mathrm{t}$ and $I_\mathrm{e}$ start oscillating after about
4\,ms. Seismometers placed on mirror mounts show that these
oscillations are mainly due to acoustic perturbations produced by
the magnet pulse and propagating from the magnet to mirror mounts
through the air. We also see that the minimum of $I_\mathrm{e}$
does not coincide with the maximum of $B^2$. This phenomenon is
due to the cavity filtering as explained in details in
Ref.\,\cite{Berceau2010}.

\begin{figure}[h]
\begin{center}
\includegraphics[width=8cm]{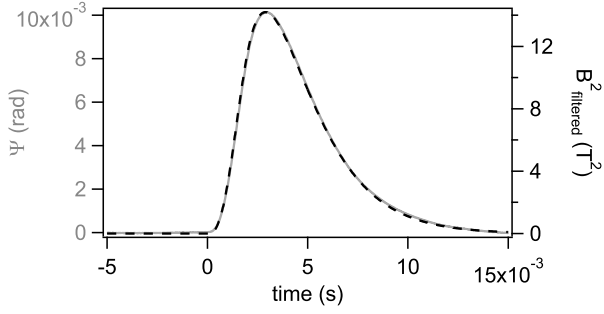}
\caption{\label{Fig:Psifiltered} Cotton-Mouton effect measurement on
32.1$\,\times\,10^{-3}$\,atm of molecular nitrogen. Gray line: Total
ellipticity as a function of time. Dashed line: Square of the
magnetic field filtered by a first-order low pass filter
corresponding to the cavity filtering.}
\end{center}
\end{figure}

In Fig.\,\ref{Fig:Psifiltered}, we plot the square of the magnetic
field filtered by the cavity and the ellipticity calculated with
Eq.\,(\ref{Eq:Psi_t}) as a function of time. Since the acoustic
perturbations affect both signals $I_\mathrm{t}$ and
$I_\mathrm{e}$, and taking into account the cavity filtering
between $I_\mathrm{t}$ and $I_\mathrm{e}$, oscillations on $\Psi$
are strongly reduced to a few $10^{-5}$\,rad, thus not visible on
this figure. These oscillations induce uncertainty to the
measurement but are already included in the A-type uncertainty on
kappa measured in section\,\ref{subsec : kappa}.

Finally, we note that both quantities $B^2_\mathrm{filtered}$ and
$\Psi$ reach their extremum at the same time and their variation
can be perfectly superimposed, providing a very precise
measurement of magnetic linear birefringence of nitrogen gas.

\subsection{Apparatus calibration}
\label{subsec : Apparatus calibration} 
\begin{figure}[h]
\begin{center}
\includegraphics[width=8cm]{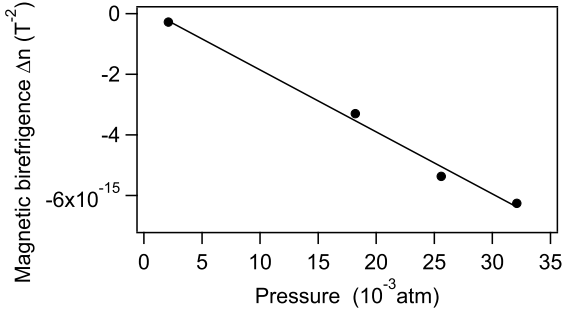}
\caption{\label{Fig:Calibration} Magnetic birefringence of
molecular nitrogen as a function of pressure. The solid line
corresponds to the linear fit of experimental data.}
\end{center}
\end{figure}

In order to calibrate our
apparatus and to evaluate its present sensitivity we have measured
the magnetic birefringence of molecular nitrogen. These
measurements have been performed at different pressure from
2.1$\,\times\,10^{-3}$ to 32.1$\,\times\,10^{-3}$\,atm and are
summarized in Fig.\,\ref{Fig:Calibration}.

In this range, nitrogen can be considered as an ideal gas and the
pressure dependence of its birefringence is thus linear:
\begin{equation}
\Delta n (\mathrm{T}^{-2})= \Delta n_\mathrm{u}(\mathrm{atm}^{-1}
\mathrm{T}^{-2})\times P (\mathrm{atm}). \label{deltanPdeltanu}
\end{equation}
We have checked that our data are correctly fitted by a linear
equation. Its $\Delta n$ axis-intercept is consistent with zero
within the uncertainties. Its slope gives the normalized magnetic
birefringence at $B = 1$\,T and $P = 1$\,atm:
\begin{equation}
\Delta n_\mathrm{u} = (-2.00 \pm 0.08 \pm 0.06)\times
10^{-13}\,\mathrm{atm}^{-1}\mathrm{T}^{-2}. \nonumber
\label{deltanN2}
\end{equation}
The first uncertainty $0.08 \times 10^{-13}$\,atm$^{-1}$T$^{-2}$
corresponds to the fitting uncertainty and represents the A-type
total uncertainty at 1$\sigma$. The second one $0.06 \times
10^{-13}$\,atm$^{-1}$T$^{-2}$ represents the B-type uncertainty at
1$\sigma$.

Our value of the normalized birefringence is compared in
Table.\,\ref{Table:N2} to other experimental published values at
$\lambda = 1064$\,nm \cite{Bregant2004,QandA}. This shows that our
value agrees perfectly well with other existing measurements. Our
total uncertainty is $ 10^{-14}$\,atm$^{-1}$T$^{-2}$, calculated
by quadratically adding the A-type and B-type uncertainties. This
is 1.8 times more precise than the other results. It therefore
provides a successful calibration of the whole apparatus.

\begin{table}[h]
\center
\begin{tabular*}{0.4\textwidth}{c c}
\hline \hline
Ref. & $\Delta n_\mathrm{u}\times10^{-13}$\\
 &  (at $P$ = 1atm and $B$ = 1T)\\
\hline
\cite{Bregant2004}    & -2.17 $\pm$ 0.21\\
  \cite{QandA}          & -2.02 $\pm$ 0.16 $\pm$ 0.08\\
    This work             &  -2.00 $\pm$  0.08 $\pm$ 0.06\\
\hline \hline
 \end{tabular*}
   \caption{\label{Table:N2} Comparison between our value of the nitrogen normalized magnetic birefringence and other experimental published values at $\lambda = 1064$\,nm.}
\end{table}

\subsection{Upper limit on vacuum magnetic birefringence measurements}
\begin{figure}[h]
\begin{center}
\includegraphics[width=8cm]{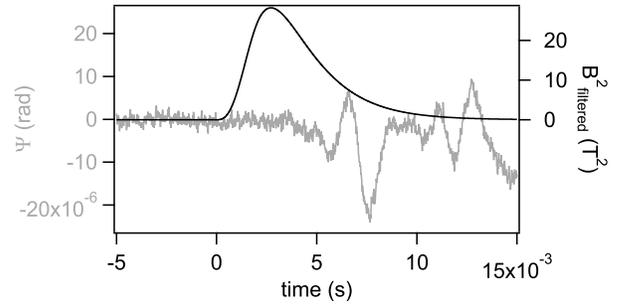}
\caption{\label{Fig:shotvacuum}Typical ellipticity (gray) measured
during a magnetic pulse (black) performed in vacuum. Acoustic
perturbations generate ellipticity oscillations starting at 4 ms.}
\end{center}
\end{figure}

Once the calibration performed we have evaluated the upper limit
of the present apparatus on vacuum magnetic birefringence. To this
end, several pulses were performed in vacuum. In
Fig.\,\ref{Fig:shotvacuum}, a typical ellipticity measured during
a magnetic pulse is plotted. Acoustic perturbations induce
oscillations of $\Psi$ starting at about 4\,ms, with variations of
the order of $10^{-5}$\,rad. In order to infer our best upper
limit for the value of the vacuum magnetic birefringence, we limit
the integration time to 4\,ms. We get $\Delta
n\,<\,5.0\,\times\,10^{-20}$\,T$^{-2}$ per pulse.

During operation, the pressure inside the UHV system was better
than $10^{-10}$\,atm. To be conservative, lets assume residual
gases are mainly 78\,\% of nitrogen and 21\,\% of oxygen. The
normalized magnetic birefringences of these gases are of the order
of $-2 \times 10^{-13}\,\mathrm{atm}^{-1}\mathrm{T}^{-2}$ and
$-2\times 10^{-12}\,\mathrm{atm}^{-1}\mathrm{T}^{-2}$ respectively
\cite{Rizzo1997}. The total residual magnetic birefringence is
then of the order of $6 \times 10^{-23}\,\mathrm{T}^{-2}$, which
is well below our current upper limit. In the final setup, vacuum
quality will be monitored with a residual gas analyzer.

\section{Conclusion}

The successful calibration we report in this paper is a crucial
step towards the measurement of the vacuum magnetic birefringence.
It shows our capability to couple intense magnetic fields with one
of the sharpest Fabry-Perot cavity in the world. It is worthwhile
to note that an energy of about 100 kJ is discharged in our coils
during a few milliseconds. These 10 MW of electrical power
generate acoustic perturbations and mechanical vibrations that
tend to misalign the cavity mirrors. The linewidth $\Delta \nu$ of
our Fabry-Perot cavity is of the order of 150 Hz. A relative
displacement $\Delta L_{c} = L_{c} \times \Delta \nu /
\nu_{\mathrm{laser}}$\,=\,1\,pm of both mirrors is enough to get
out of resonance. The sharper the cavity, the bigger the
challenge.

The sensitivity per pulse we got both in gases and in vacuum is
the best ever reached for this kind of measurement. For sake of
comparison, the best birefringence limit obtained in vacuum with
continuous magnets is $\Delta n\,\leq\,2.1\,\times\,10^{-20}$
$\mathrm{T}^{-2}$ with an integration time of $t_{\mathrm{int}}$ =
65\,200\,s \cite{PVLAS2008}. In order to compare both methods, we
need to translate the best limit obtained in continuous regime to
the one obtained with our integration time
$T_\mathrm{i}$\,=\,4\,ms. Assuming white noise for both methods,
best limit reported in Ref.\,\cite{PVLAS2008} corresponds to
$\Delta n(T_\mathrm{i}) = \Delta
n(t_\mathrm{int})\sqrt{t_\mathrm{int}/T_\mathrm{i}}\leq\
8.5\,\times\,10^{-17}$ in 4\,ms of integration. This value is more
than three orders of magnitude higher than ours, proving that
pulsed fields are a powerful tool for magnetic birefringence
measurements.

Long terms perspective is to get a value of $\Delta n =
4\,\times\,10^{-24}\mathrm{T}^{-2}$, corresponding to the vacuum
magnetic birefringence, with at most 1000 pulses. This corresponds
to a sensitivity better than
$1.3\,\times\,10^{-22}\mathrm{T}^{-2}$ per pulse. A factor of the
order of 10 on optical sensitivity will be achievable with a
better acoustic insulation and a more robust locking system,
especially reducing the noise of the measured light intensities
transmitted by the cavity. Further improvements depend on the
possibility to have higher magnetic fields. We have designed a new
pulsed coil, called XXL-coil, which has already reached a field
higher than 30\,T when a current higher than 27\,000\,A is
injected. This corresponds to more than 300\,T$^2$m
\cite{sitewebLNCMI}. Two XXL-coils will allow us to improve our
current sensitivity by a factor 100. In the near future, the
apparatus will be modified in order to host these XXL-coils.
Therefore the final version of the experiment will be ready for
operation.

\begin{acknowledgments}

We thank all the members of the BMV collaboration, and in particular
J. B\'eard, J. Billette, P. Frings, J. Mauchain, M. Nardone, L.
Recoules and G. Rikken for strong support. We are also indebted to
the whole technical staff of LNCMI. We acknowledge the support of
the \textit{Fondation pour la recherche IXCORE} and of the {\it
ANR-Programme non th\'{e}matique} (ANR-BLAN06-3-139634).

\end{acknowledgments}

\bibliography{apssamp}

\begin{thebibliography}{}
\bibitem{It_Ie_justification} Fig.\,\ref{Fig:finesse_Ie} and Fig.\,\ref{Fig:finesse_ATF} do not correspond to the same run of data and thus can't be directly compared.
\end{thebibliography}

\end{document}